%% file: sample62.tex
\documentclass[twocolumn]{aastex62}

\submitjournal{ApJ}

\shorttitle{Wide brown dwarf binary}
\shortauthors{Faherty et al.}

\begin{document}

\title{WISE2150-7520AB: A very low mass, wide co-moving brown dwarf system discovered through the citizen science project Backyard Worlds: Planet 9.\footnote{This paper includes data gathered with the 6.5 meter Magellan Telescopes located at Las Campanas Observatory, Chile.}}

\correspondingauthor{J. Faherty}
\email{jfaherty@amnh.org}

\author[0000-0001-6251-0573]{Jacqueline K. Faherty}
\affil{Department of Astrophysics, American Museum of Natural History, Central Park West at 79th Street, NY 10024, USA }

\author[0000-0003-2236-2320]{Sam Goodman}
\affiliation{Backyard Worlds: Planet 9}

\author[0000-0001-7896-5791]{Dan Caselden}
\affiliation{Backyard Worlds: Planet 9}

\author[0000-0002-7630-1243]{Guillaume Colin}
\affiliation{Backyard Worlds: Planet 9}

\author[0000-0002-2387-5489]{Marc J. Kuchner}
\affiliation{NASA Goddard Space Flight Center, Exoplanets and Stellar Astrophysics Laboratory, Code 667, Greenbelt, MD 20771}

\author[0000-0002-1125-7384]{Aaron M. Meisner}
\affiliation{National Optical Astronomy Observatory, 950 N. Cherry Ave., Tucson, AZ
85719, USA}
\altaffiliation{Hubble Fellow}

\author[0000-0002-2592-9612]{Jonathan Gagn\'e}
\affiliation{Institute  for  Research  on  Exoplanets, Université  de  Montréal, 2900  Boulevard  Édouard-Montpetit Montréal,  QC  Canada  H3T  1J4}

\author[0000-0001-5106-1207]{Adam C. Schneider}
\affiliation{School of Earth and Space Exploration, Arizona State University, Tempe, AZ, 85282, USA}

\author[0000-0003-4636-6676]{Eileen C. Gonzales}
\affiliation{Department of Astrophysics, American Museum of Natural History, Central Park West at 79th Street, NY 10024, USA }
\affiliation{The Graduate Center, City University of New York, New York, NY 10016, USA}
\affiliation{Department of Physics and Astronomy, Hunter College, City University of New York, New York, NY 10065, USA}

\author[0000-0001-8170-7072]{Daniella C. Bardalez Gagliuffi}
\affiliation{Department of Astrophysics, American Museum of Natural History, Central Park West at 79th Street, NY 10024, USA }

\author[0000-0002-9632-9382]{Sarah E. Logsdon}
\affiliation{NASA Goddard Space Flight Center, Exoplanets and Stellar Astrophysics Laboratory, Code 667, Greenbelt, MD 20771}
\altaffiliation{NASA Postdoctoral Program Fellow}

\author{Katelyn Allers}
\affiliation{Physics and Astronomy Department,
Bucknell University,
701 Moore Ave, Lewisburg, PA 17837
Norman, OK 73019}

\author{Adam J. Burgasser}
\email{aburgasser@ucsd.edu}
\affiliation{Department of Physics, Center for Astrophysics and Space Sciences, Mail Code 0424, 9500 Gilman Drive, La Jolla, CA 92093-0424 USA}

\author{The Backyard Worlds: Planet 9 Collaboration}
\affiliation{Backyard Worlds: Planet 9}
 \begin{abstract}
We report the discovery of WISE2150-7520AB (W2150AB): a widely separated ($\sim$ 341 AU) very low mass L1 + T8 co-moving system. 
The system consists of the previously known L1 primary 2MASS J21501592-7520367 and a newly discovered T8 secondary found at position 21:50:18.99 -75:20:54.6 (MJD=57947) using Wide-field Infrared Survey Explorer (WISE) data via the Backyard Worlds: Planet 9 citizen science project. We present {\it Spitzer} $ch1$ and $ch2$ photometry ($ch1$-$ch2$= 1.41 $\pm$0.04 mag) of the secondary and FIRE prism spectra of both components. The sources show no peculiar spectral or photometric signatures indicating that each component is likely field age. Using all observed data and the Gaia DR2 parallax of 41.3593 $\pm$ 0.2799 mas for W2150A we deduce fundamental parameters of log(L$_{bol}$/L$_{\sun}$)=-3.69 $\pm$ 0.01,  $T_{\rm eff}$=2118 $\pm$ 62 K, and an estimated mass=72 $\pm$ 12 $M_{\rm Jup}$ for the L1 and log(L$_{bol}$/L$_{\sun}$)=-5.64 $\pm$ 0.02, $T_{\rm eff}$=719 $\pm$ 61 K, and an estimated mass=34 $\pm$ 22 $M_{\rm Jup}$ for the T8. At a physical separation of $\sim$341 AU this system has E$_{bin}$ = 10$^{41}$ erg making it the lowest binding energy system of any pair with M$_{tot}$ $<$ 0.1 M$_{Sun}$ not associated with a young cluster.  It is equivalent in estimated mass ratio, E$_{bin}$, and physical separation to the $\sim$ 2 Myr M7.25 + M8.25 binary brown dwarf 2MASS J11011926-7732383AB (2M1101AB) found in the Chameleon star forming region. W2150AB is the widest companion system yet observed in the field where the primary is an L dwarf or later.

\end{abstract}


\section{Introduction} \label{sec:intro}
Brown dwarfs are a unique population of astronomical objects and a critical bridge between stars and planets.  On the high mass end, brown dwarfs overlap in observable properties with the coolest stars like TRAPPIST-1 which hosts seven terrestrial worlds (\citealt{Gillon17}).  On the low mass end, brown dwarfs overlap with the observable properties of directly imaged exoplanets like 51 Eri b (\citealt{Macintosh15}) and Beta Pictoris b (\citealt{Lagrange10}).  On the coolest end brown dwarfs like J085510.83-071442.5 -- a $\sim$ 250 K object at just 2 pc from the Sun (\citealt{Luhman14}) -- are directly comparable to Jupiter (\citealt{Skemer16}, \citealt{Morley18}).

Studying brown dwarfs provides insight into stellar and planetary atmospheres and activity.  One of the most important and outstanding questions in substellar mass science is how these objects form and evolve.   Co-moving companions are a key sub-population for investigating questions of formation.

Early searches for low mass companions resulted in two distinct categories of objects.  Those that were either (1) well-resolved companions discovered through common proper motion or closely separated and with statistically consistent distances (e.g. \citealt{Kirkpatrick01},\citealt{Wilson01}, \citealt{Faherty10}) or those that were (2) closely bound and discovered through high resolution imaging (e.g. \citealt{Martin99}, \citealt{Koerner99}, \citealt{Burgasser03B}).  For several years the only objects that fell in category (1) were brown dwarfs orbiting higher mass stars (mass ratios $<<$ 1) and those that fell in category (2) were near equal mass binaries (mass ratio $\sim$ 1) with very low total masses ($\sim$ 0.1 M$_{\sun}$) and binding energies.

Work done on young clusters such as Taurus, Rho Ophiucus, and Chameleon resulted in the discovery of widely separated objects ($>$ 100 AU) with mass ratios near 1 and small total masses that were hybrids between the two previously distinct classes (e.g. \citealt{Luhman04}, \citealt{Close07}).  Searches in the field also turned up a handful of objects that were widely separated with relatively low total masses, though nothing that rivaled the low binding energies found among young cluster companions (e.g. \citealt{Artigau07}, \citealt{Radigan09}). 

Brown dwarf spectral classification categories include L, T, and Y dwarfs (e.g. \citealt{Kirkpatrick99}, \citealt{Burgasser06}, \citealt{Cushing11} -- M dwarfs are almost exclusively stellar unless young). In the absence of dynamical mass measurements, an age is required to determine whether an object of a given spectral class has a mass that is above or below the Hydrogen burning limit.  However the L, T, and Y classes are all definitively in the low temperature regime for compact sources (e.g. \citealt{Vrba04}, \citealt{Dupuy13} \citealt{Tinney14}, \citealt{Filippazzo15}).  Consequently, the bulk of ultracool dwarf companions (e.g. L+L,L+T, T+T, T+Y) have been found in category (2): closely bound and unresolved in all but high resolution imaging.  Until this work there were only two easily resolved visual L+T binaries:  SDSS J1416+13AB (\citealt{Burningham10}) which is an L7+T7.5 binary with an angular separation of 9.81$\arcsec$ or a physical separation of 89.3$\pm$1.5~AU at the system's distance of 9.1$\pm$0.15pc (\citealt{Dupuy12}) and Luhman 16AB (\citealt{Luhman13}) which is an L7.5+T0.5 binary with an angular separation of 1.5$\arcsec$ or a physical separation of $\sim$3~AU at the systems distance of 2.02$\pm$0.019.

Brown dwarf formation theories are specific in their predictions of binary parameters. Model scenarios that involve ejection (e.g. \citealt{Bate02, Bate11}; \citealt{Reipurth01}), turbulent fragmentation (e.g. \citealt{Padoan04}), and/or disc fragmentation (e.g. \citealt{Goodwin07}; \citealt{Li15}) to produce brown dwarfs predict statistical properties which can be compared to observational studies as evidence for or against formation pathways.  
In general, theoretical models do not produce very low mass binaries (M$_{tot}$ $<$ 0.1 M$_{Sun}$) with separations $>$ 10 AU (e.g. \citealt{Bate02}) that can survive to field age. This prediction is roughly consistent with the very low mass wide binary population of near equal mass companions. The exceptions are those found in clusters and the handful of slightly higher mass wide objects. 
This dearth of observed widely separated companions was explained because they were thought to either not exist or not survive to field age. 

In this paper we report the discovery of a wide very low mass co-moving system consisting of an L1 and T8 discovered through the citizen science project Backyard Worlds:  Planet 9. Section 2 reviews how the discovery was made. Section 3 describes new data acquired on the primary and secondary sources in the system. Section 4 has observational details on each component. Section 5 discusses the $Gaia$ parallax, kinematics of each component and probability of chance alignment.  Section 6 has the color magnitude diagram analysis for the system while section 7 reviews the age.  Section 8 details the fundamental parameters for each component and section 9 has the binding energy analysis.  Conclusions are summarized in section 10. 

\section{Discovery}
The Backyard Worlds: Planet 9 citizen science project (Backyard Worlds for short) has been operational since February 2017. The scientific goal of the project is to complete the census of the solar neighborhood (including the solar system, e.g. Planet 9) with objects that are detectable primarily at mid infrared wavelengths and that were missed by previous searches (see \citealt{Kuchner17} and \citealt{Debes19}). Backyard Worlds utilizes multiple epochs of NASA's Wide-field Infrared Survey Explorer (WISE) mission at both W1 ($\sim$3.5 $\mu$m) and W2 ($\sim$4.5$\mu$m) wavelengths. Project participants are asked to blink between four unWISE images (see \citealt{Meisner17b}) where the time-span between the first and last image is $\sim$ 4.5 years. Given this time baseline, objects of significant motion (e.g. $>$ 200 mas yr$^{-1}$) are relatively easy to visually identify (see for example \citealt{Kuchner17}). 

The BackyardWorlds.org website hosted by Zooniverse provides two avenues for reporting a proper motion candidate of scientific interest to the Backyard Worlds: Planet 9 research team.  The first is by using the web portal to blink WISE epoch images and ``click" on an object that appears to move over the $\sim$4.5 year baseline.  Once identified, these objects go into a large repository that the research team can access through Zooniverse. The second avenue for reporting a candidate is to alert the science team by submitting the coordinates and details of the source on a Google form labeled ``Think You've Got One".

Three citizen science users (Co-authors S. Goodman, D. Caselden, and G. Colin) brought to our attention a WISE $W2$ only detected source with significant motion. They used the Google form and emphasized the objects importance by emailing the Backyard Worlds distribution list as well as key researchers on our team. In addition, these users easily noted a bright source $\sim$14.1 $\arcsec$ away that appeared to be co-moving.  Upon further investigation the users realized this was the known L1 dwarf SIPS J2150-7520 (or source 2MASS J21501592-7520367; \citealt{Deacon2005}).   On 29 June 2018, the motion of W2150B was vetted by the research team and added to our high priority follow-up target list.  Figure~\ref{fig:finder} shows a screenshot from the WISEview website (\citealt{Caselden18}) which was used to identify and confirm the system.

\section{Data}
We obtained both near infrared spectra and mid infrared photometry for the system WISE2150-7520AB (W2150AB).  

\subsection{Magellan FIRE Spectroscopy}
We used the 6.5m Baade Magellan telescope and the Folded-port InfraRed Echellette (FIRE; \citealt{Simcoe13}) spectrograph to obtain near-infrared spectra of both the primary and the secondary in this system.  Observations were made on 01 December 2018 under clear conditions.  For all observations, we used the prism mode and the 0.6$\arcsec$ slit  (resolution $\lambda$/$\Delta \lambda \sim$ 100) covering the full 0.8 - 2.5 $\micron$ band. We observed the A component using a standard ABBA nod pattern with an exposure time of 90 seconds per nod. For the B component we obtained an ABBAAB nod pattern with an exposure time of 120 seconds per nod.  Immediately after the B component, we observed the A star HD 200523 for telluric correction and obtained a Ne Ar lamp spectrum for wavelength calibration.  At the start of the night we used quartz lamps as domeflats in order to calibrate pixel-to-pixel response.  Data were reduced using the FIREHOSE package which is based on the MASE and SpeXtool reduction packages (\citealt{Bochanski09}, \citealt{Cushing04}, \citealt{vacca03}). 

\subsection{Spitzer Photometry}
The field surrounding W2150B was observed by the Spitzer Space telescope on 2018-12-07.  Data were obtained in both $ch1$ and $ch2$ bands.  Peak-up was disabled and each filter was observed using a 16 position spiral dither pattern with 30 seconds per frame.  The readout was done in full array mode.  Data were downloaded and aperture photometry was performed on the Spitzer Heritage reduced mosaic images.  

\begin{figure}[ht!]
\begin{center}
\includegraphics[width=1.0\linewidth]{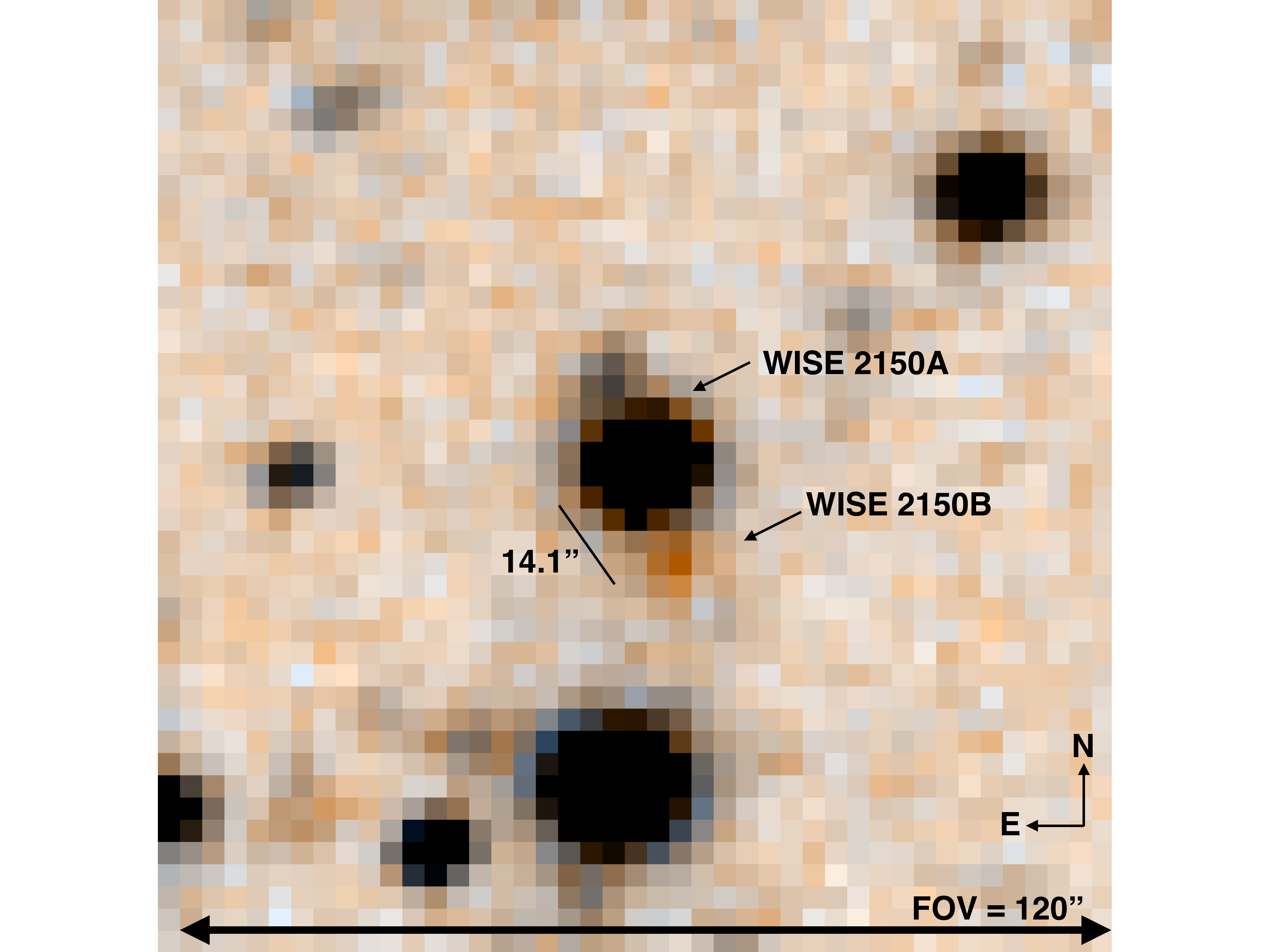}
\caption{The finder chart for the W2150AB system taken from the WISEVIEW website (\citealt{Caselden18}).   To see the animated motion between available WISE epochs visit the URL byw.tools/wiseview and use coordinates RA,DEC=327.576919, -75.34805934. The color choice combines WISE W1 and W2 images where ''orange" sources are strong W2 and weak W1 detections. }
\label{fig:finder}
\end{center}
\end{figure}

\begin{figure}[ht!]
\begin{center}
\includegraphics[width=1.0\linewidth]{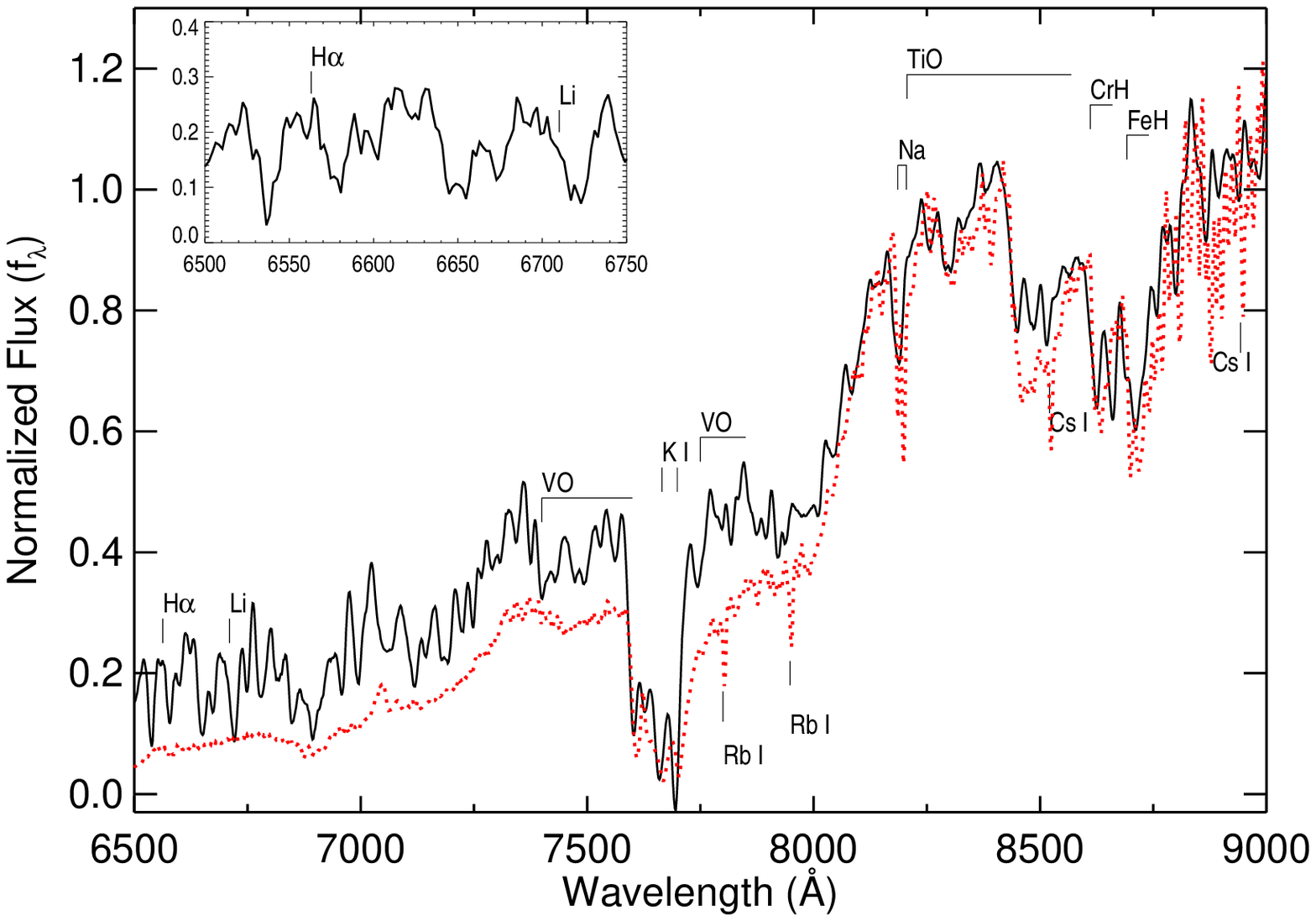}
\includegraphics[width=1.0\linewidth]{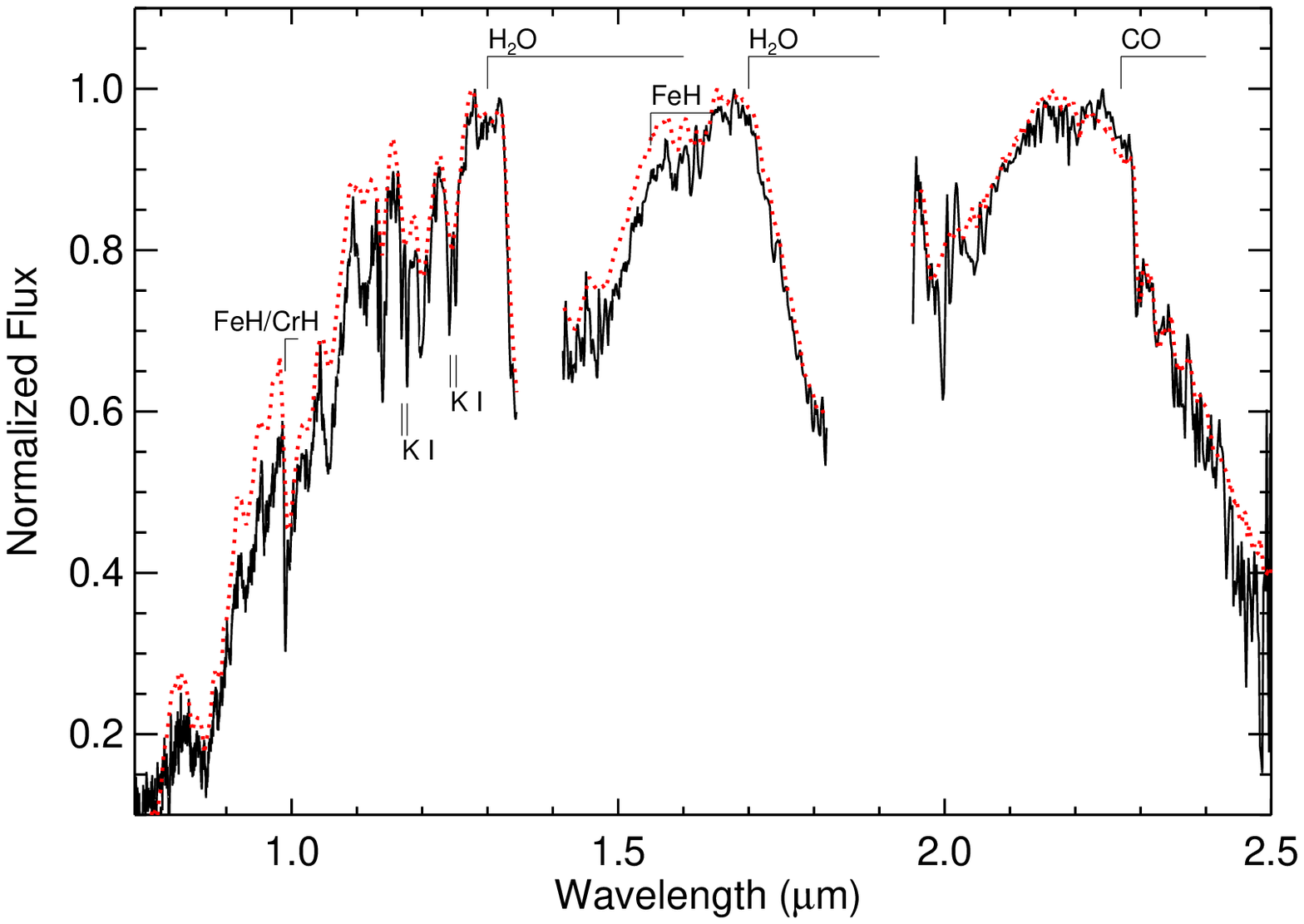}
\caption{Top: The optical spectrum of W2150A (black solid line) from \citet{Reid2008} compared to the L1 standard 2MASSW J2130446-084520 (red dashed; also from \citealt{Reid2008}).  Prominent spectral absorption features are highlighted.  The region surrounding the H$\alpha$ and Li I features is contained in the inset; we find no detection of either.  Bottom:  The infrared spectrum of W2150A (black solid line) obtained using the FIRE spectrograph in prism mode (this work), normalized to the peak in each band (JHK). We compare this spectrum to that of the L1 standard 2MASSW J2130446-084520 (red dashed curve) from \citealt{Kirkpatrick10}.  Prominent near infrared features are labeled.}
\label{fig:W2150A-Spectrum}
\end{center}
\end{figure}

\begin{figure}[ht!]
\begin{center}
\includegraphics[width=1.0\linewidth]{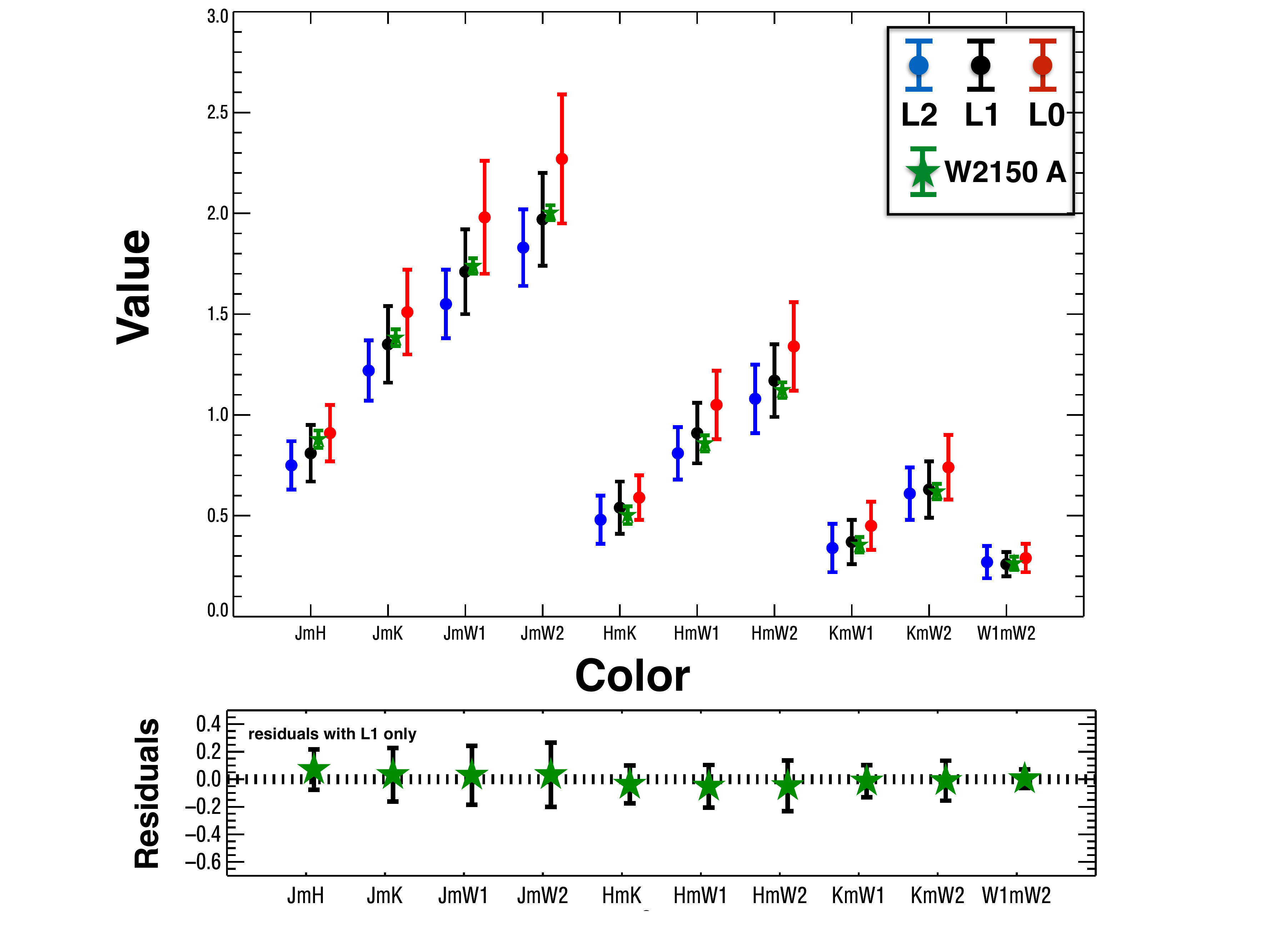}
\includegraphics[width=1.0\linewidth]{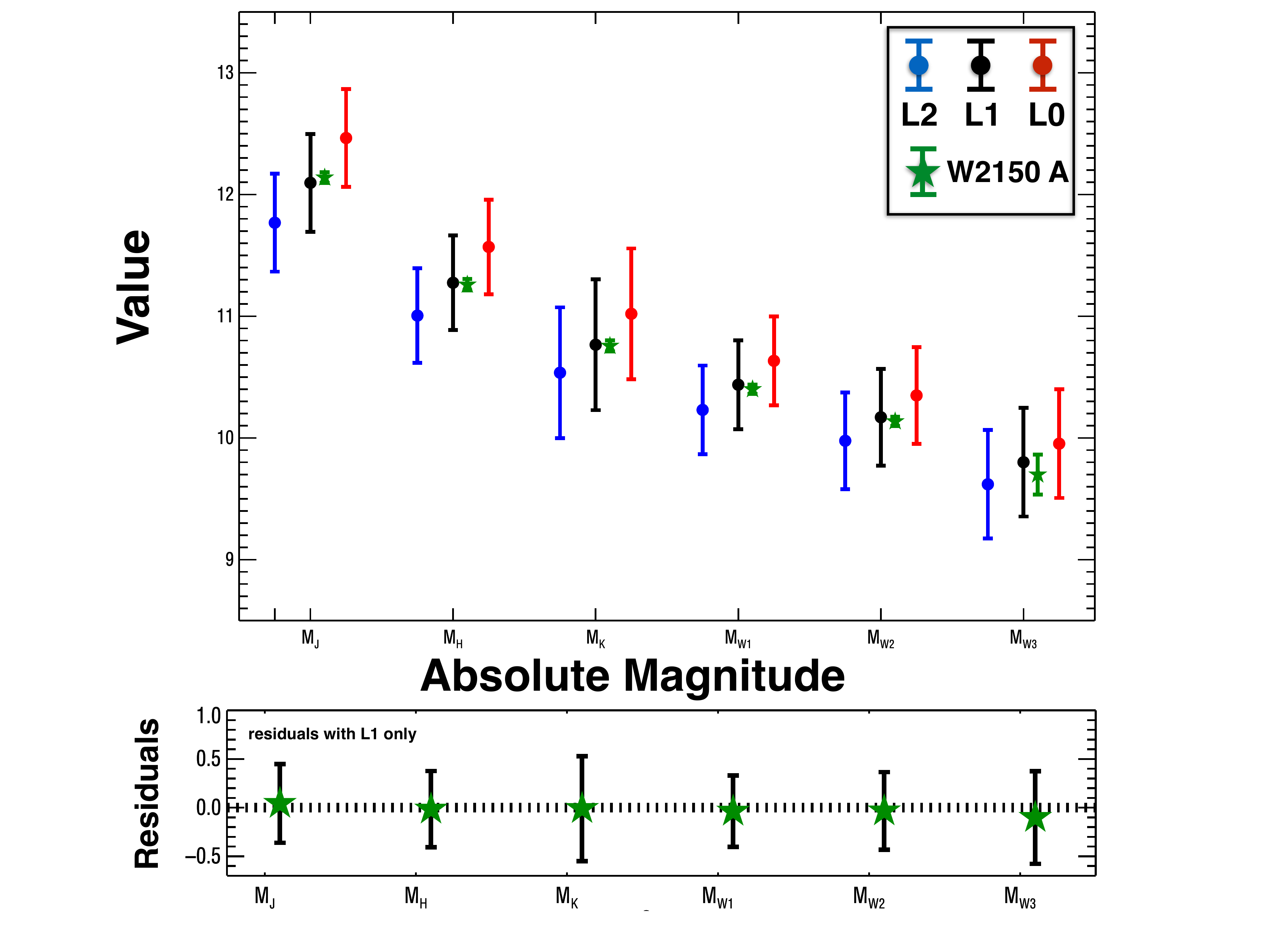}
\caption{Top: Average colors and their spread for  2MASS $JHK$ through WISE $W1W2$ bandpasses for field L0 (red), L1 (black), and L2 (blue) dwarfs as listed in \citet{Faherty16}. We show the values calculated for W2150A with uncertainties as a five point star colored in green.  Bottom:  Average absolute magnitudes and their spread as above.  The values for W2150A are calculated using $2MASS$ and WISE photometry as well as the Gaia DR2 parallax.}
\label{fig:Colors-AbsMag-2150A}
\end{center}
\end{figure}
\section{Details on the components}
\subsection{Primary}
W2150A was originally reported as a proper motion source with red optical and near infrared colors reminiscent of a late-type M/early L dwarf in \citealt{Deacon2005} (originally called SIPS J2150-7520 and identified in 2MASS as 2MASS J21501592-7520367).  It was followed up with optical spectroscopy by \citet{Reid2008} and published as an L1$\pm$1.  Subsequently, the proper motion was updated for this object in \citet{Casewell2008} and \citet{Faherty09}.  \citet{Faherty10} specifically looked for co-moving companions to known M and L dwarfs but nothing of note was recovered around W2150A in either the Hipparcos or LSPM catalogs. 

In Figure ~\ref{fig:W2150A-Spectrum} we plot the optical spectrum from \citet{Reid2008} and the near infrared spectrum obtained with FIRE highlighting prominent spectral features. Overplotted on each is the L1 near infrared standard 2MASSW J2130446-084520.  The sources are normalized by the maximum flux over the wavelength covered (J,H and K). In the optical, H$\alpha$ and Li I absorption are indicators of activity as well as mass (hence age). The optical spectrum of W2150A shows no detectable Li I or H$\alpha$ absorption or emission although the noise is significant. There is also no Rb I or appreciable Cs I detected although this may be due to the low resolution of the data.  The near infrared spectrum receives a field gravity designation using the \citet{Allers13} spectral indices and we see no visible signatures of low surface gravity.   

\begin{figure}[ht!]
\begin{center}
\includegraphics[width=1.0\linewidth]{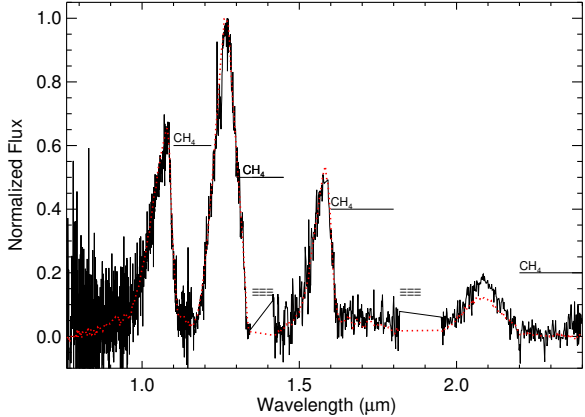}
\caption{The FIRE prism spectrum of W2150B (black curve) compared to a spectrum of the T8 standard 2MASSI J0415195-093506 (red curve) from \citealt{Burgasser04}.  The spectra has been normalized by the maximum flux over all the data. }
\label{fig:Spectrum-2150B}
\end{center}
\end{figure}

Figure ~\ref{fig:Colors-AbsMag-2150A} shows the colors and absolute magnitudes from $2MASS$ through WISE for W2150A compared to median values of field objects as listed in \citet{Faherty16}. W2150A shows no deviation and fits within 1$\sigma$ of seemingly normal equivalent type objects.  All positional, photometric, and kinematic data are listed in Table ~\ref{tab:param}.   

\begin{figure*}
\begin{center}
\includegraphics[width=1.0\linewidth]{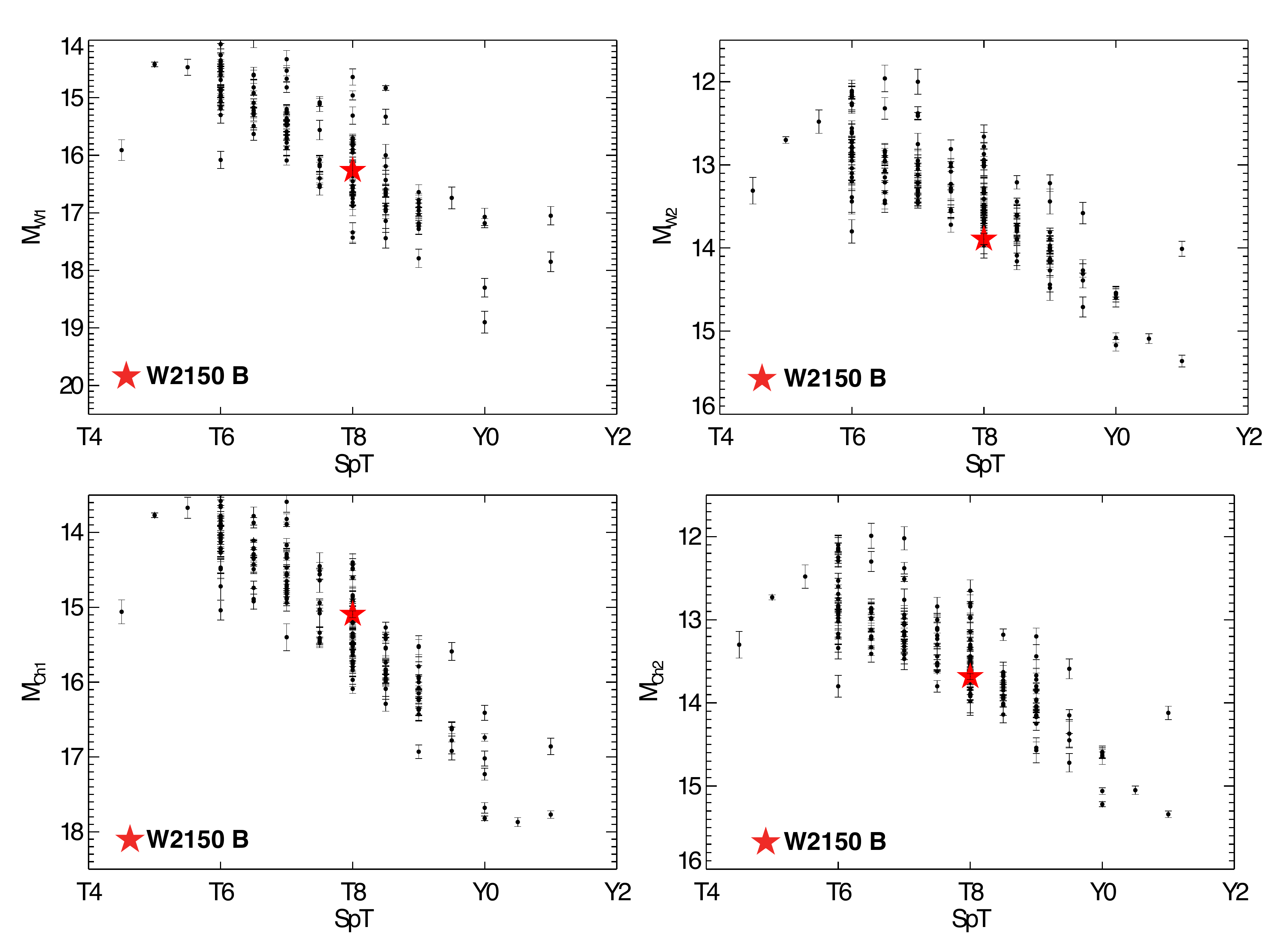}
\caption{Absolute magnitude versus spectral type diagram for W2150B in the WISE and Spitzer bandpasses.  The mid to late-type T comparative sample comes from \citet{Kirkpatrick19} with a few earlier type sources from \citet{Dupuy12}.}
\label{fig:CMD-SpT}
\end{center}
\end{figure*}

\begin{figure*}
\begin{center}
\includegraphics[width=1.0\linewidth]{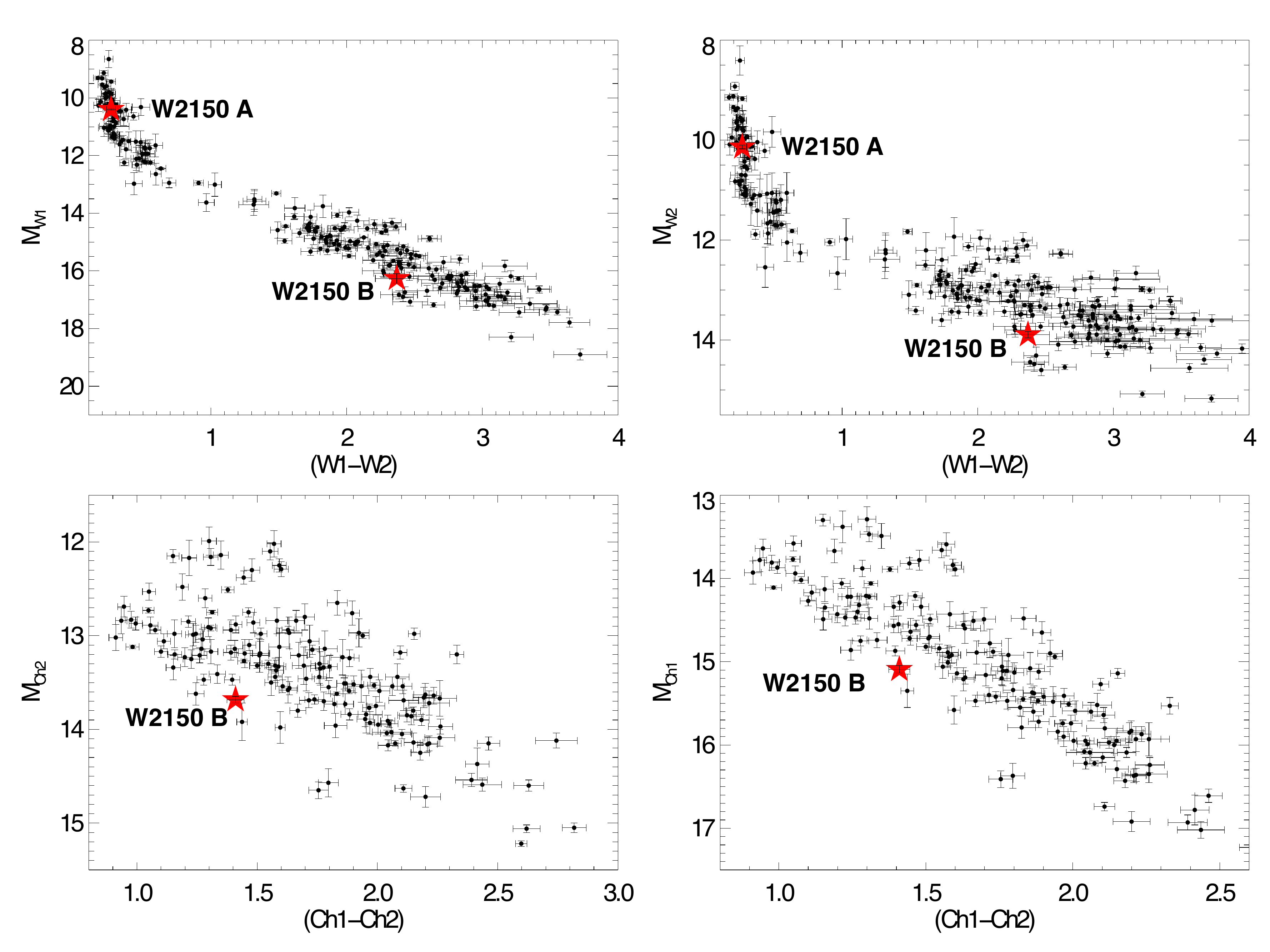}
\caption{Color magnitude diagrams for L through late-type T/Y dwarfs with W2150AB highlighted. The top panels show WISE color magnitude diagrams and both the primary and secondary.  The bottom panels show {\it Spitzer} color magnitude diagrams, including only the secondary. The mid to late-type T dwarfs in the comparative sample come from \citet{Kirkpatrick19} and the earlier type objects come from \citet{Dupuy12} and \citet{Faherty12}.}  
\label{fig:CMD}
\end{center}
\end{figure*}

\subsection{Secondary\label{sec:secondary}}
At the time of its initial discovery, W2150B had not been detected in any previously published WISE catalog (for example it was in neither AllWISE nor the WISE All-Sky catalog) nor any publicly accessible catalog (e.g. it was also not in 2MASS).  After citizen scientists reported the source at position 21:50:18.99 -75:20:54.6 (MJD=57947), the research team checked the unWISE Catalog for a detection (\citealt{Lang14}; \citealt{Meisner17}; \citealt{Schlafly19}) and found that the source was faint (W2=16.01$\pm$0.06) and red in the WISE bands ((W1-W2)=2.42$\pm$0.15). This red color, combined with the source's significant motion, were a tell-tale sign that the object was a close and cold brown dwarf.  

The unWISE Catalog provides valuable information about W2150B, but the photometry in this catalog was derived from the full 2010-2017 unWISE stack, with no attempt made to account for the considerable $\sim$7$''$ of source motion during that time span. We therefore performed additional, custom WISE astrometry and photometry using coadds binned into yearly time intervals \citep{Meisner18}. We bin into yearly intervals because W2150B is so faint that coaddition of two six-month WISE sky passes is necessary to form one measurement epoch. The field surrounding W2150B was observed by WISE in two separate sky passes during 2010 (mean MJD=55402), 2015 (mean MJD=57198), 2016 (mean MJD=57580), and 2017 (mean MJD=57947).  The field should have also been observed by WISE in April 2014, but due to a command timing anomaly, the location suffered a missed sky pass.
We therefore omit calendar year 2014 WISE data from our analysis. Combining measurements from the remaining four coadds yields $W2$=15.81$\pm$0.05 mag.  This value agrees well with the $W2$ estimate reported in the unWISE Catalog, with the latter being slightly faint due to its smeared point spread function. The source is marginally detected in $W1$ yearly coadds and therefore we did not attempt to re-do photometry in this band.  Instead we adopt $W1$ = 18.18$\pm$0.15, which is the unWISE Catalog mag with a correction to account for the smearing due to the high proper motion of the source. W2150B was observed in Spitzer GO program 14076 and both $ch1$ and $ch2$ photometry were acquired to verify the cold nature of the source. We measured a ($ch1$-$ch2$) color for the object of 1.41$\pm$0.04 mag.  According to the \citet{Kirkpatrick19} color spectral type relations, the ($ch1$-$ch2$) and (W1-W2) colors correspond to a spectral type of T7$\pm$1. 

The FIRE prism spectrum for W2150B is shown in Figure ~\ref{fig:Spectrum-2150B} and appears most like a field T8.  The source matches the T8 infrared standard well, with the exception of enhanced flux seen in the K band. 

Using the Gaia parallax of the primary, we computed absolute magnitudes in WISE $W1W2$ and {\it Spitzer} $ch1ch2$ bands.  Figure ~\ref{fig:CMD-SpT} shows a suite of absolute magnitude versus spectral type diagrams featuring the WISE and {\it Spitzer} bandpasses discussed in this work.  W2150B falls within the spread of normal field T8 objects discussed in \citet{Kirkpatrick19}. 

Figure~\ref{fig:CMD} shows the color magnitude diagrams for brown dwarfs in WISE and {\it Spitzer} bands.  Assuming the Gaia DR2 parallax for W2150B (see section~\ref{sec:CMD}), its absolute magnitudes in each band are on the faint side for its color.   It remains unclear if such a position on color magnitude diagrams might indicate slightly deviant cloud, metallicity or gravity properties (e.g. \citealt{Tinney14}, \citealt{Leggett17}). 

In summary, W2150B appears to be a spectrally normal field T8 dwarf.  It is well matched to the absolute magnitudes of known similar type objects yet slightly faint for its color in all bands. All positional, photometric, and kinematic data are listed in Table ~\ref{tab:param}.

\section{Gaia with probability of chance alignment\label{sec:chancealign}} Using the WISEVIEW motion visualization tool (\citealt{Caselden18})\footnote{D. Caselden is also one of the citizen scientist co-discoverers of this binary.}, the motions of both the primary and secondary are obvious (see Figure ~\ref{fig:finder} for a screenshot).  After cross-matching with the Gaia DR2 catalog release (\citealt{GaiaDR2}, \citealt{GaiaCollaboration2016}) citizen scientist and co-author S. Goodman reported that the primary was also source Gaia DR2 6358287868675805824 and had a well measured parallax and proper motion.  The Gaia parameters for 2M2150A -- which we assume as the astrometry for the system -- are listed in Table~\ref{tab:param}.  

While the WISEVIEW animation clearly shows the two sources are moving at a similar rate, we computed the proper motion for W2150B by examining the yearly WISE coadds of $\S$\ref{sec:secondary} in combination with the Spitzer position. As we mentioned in  $\S$\ref{sec:secondary}, there are four yearly WISE coadds for W2150B with mean MJDs of 55402, 57198, 57580, and 57947. Including the Spitzer image taken at MJD=58459 provides an 8.37 year baseline between the first and last position measurements.  We calculated ($\mu_{ra}$cos(dec), $\mu_{dec}$)=(876$\pm$45, $-$278$\pm$45) mas yr$^{-1}$ and found that our calculated proper motions in both RA and DEC for W2150B are within 1$\sigma$ of the Gaia DR2 values for W2150A.  All kinematic information for W2150B is listed in Table~\ref{tab:param}. 

To quantify the probability that the system might be a chance alignment, we examined the 100pc Gaia DR2 catalog and found all the objects with proper motion component and parallax values that fell within 1$\sigma$ of W2150B.  Out of the 700,055 stars there were 4 matches including W2150A.  W2150A is $<  15 \arcsec$ away while the three other matches were scattered across the sky (hundreds of degrees away).  We ran a Monte Carlo simulation with 90,000 iterations of randomly moving stars to determine that there was a 0.00007\% likelihood that W2150A is a chance coincidence with W2150B (at an angular separation of $<$ 15 $\arcsec$). 

Furthermore, the Gaia DR2 parallax for W2150A matches within 1$\sigma$ with the estimated spectrophotometric distance for W2150B that comes from the spectral type relations in \citet{Kirkpatrick19} (see section ~\ref{sec:secondary} and Table ~\ref{tab:param}). 

\section{Color Magnitude Diagrams for the System\label{sec:CMD}} For all analysis that follows, we assume that W2150AB is a physically associated, co-evolving system; therefore we use the Gaia parallax of the primary for the secondary as well.  We examined both sources on color magnitude diagrams to investigate commonality or differences in their positions relative to the field sequence.  The top panels of Figure ~\ref{fig:CMD} show the WISE color magnitude diagrams for field L through late-type T dwarfs with both W2150A and W2150B highlighted.  While W2150A sits well within the locus of field sources, W2150B is faint and/or blue compared to equivalent sources.  Given that we also had {\it Spitzer} photometry for the secondary and there is an array of trigonometric parallax and photometric data on late-type sources from the \citet{Kirkpatrick19} program, we also show the {\it Spitzer} color magnitude diagrams in the bottom panels of Figure ~\ref{fig:CMD}.  Similar to the WISE diagrams, W2150 B appears slightly faint and/or blue for its given color compared to the field sequence.  This position suggests that the source is lower metallicity and hence older than the field population (e.g. \citealt{Leggett10}) however nothing conclusive can be drawn at this time.  The primary shows no indications of high surface gravity and none of the classic low metallicity/subdwarf spectral indicators (e.g. \citealt{Kirkpatrick14, Kirkpatrick16}, \citealt{Burgasser04}, \citealt{Gonzales18}, \citealt{Cushing09}).  

\section{Discussion on the Age of the System\label{sec:age}}
Identifying the age of a brown dwarf is an extremely difficult task.  Solar type stars have a suite of diagnostics to estimate the age range of an object such as gyrochronology, astroseismology, chromospheric activity, and Li I depletion (see for e.g \citealt{Barnes07}, \citealt{Mamajek08}, \citealt{Pavlenko96}). However brown dwarfs have relatively few diagnostic tools largely due to their cool temperatures and lack of stable core nuclear burning.  There are a handful of emerging diagnostics but they are not precisely calibrated.  For instance, spectral features like alkali lines, metal oxide and hydride bands and overall H band shape can strongly indicate whether an object is low/high surface gravity hence young/old (see for example \citealt{Allers13}, \citealt{Cruz09}).  Near infrared colors combined with spectral type and kinematics can also indicate sub populations of brown dwarfs which are redder/slower/younger or bluer/faster/older.  While there are now several L and T dwarfs known in nearby moving groups (e.g. \citealt{Faherty16, Faherty13}, \citealt{Gagne17, Gagne18}, \citealt{Artigau15}, \citealt{Liu13, Liu16}, \citealt{Riedel19}) as well as substellar mass subdwarfs (e.g. \citealt{Burgasser03}, \citealt{Kirkpatrick14}) age diagnostics show significant scatter and are not established for field age objects.  

Co-moving systems are excellent sources of calibration for age diagnostics.  For this purpose, one would like to find a brown dwarf co-moving with a higher mass companion (solar type star for instance), take an age diagnostic for the primary and apply it to the secondary to calibrate its observable features (see, e.g. \citealt{Kirkpatrick10}, \citealt{Faherty10,Faherty11}, \citealt{Deacon14}, \citealt{Burningham13}, \citealt{Muzic12}).  The case of W2150AB is intriguing because it is two well resolved ultracool dwarf objects orbiting each other.  By examining their spectral features and positions on color magnitude and spectral photometric diagrams in tandem we can investigate whether there are any common age-diagnostic trends in their values.  However, as noted above, neither object shows peculiar spectral features.  While W2150B appears slightly faint for its given WISE or Spitzer color, neither it nor W2150A are in a particularly extreme part of any diagram.  

The total tangential velocity of the system is $>$ 100 km s$^{-1}$ (as shown in table ~\ref{tab:param}), which is notable especially given the position of W2150B on Figure ~\ref{fig:CMD}.  \citet{Faherty09} found that objects with vtan $>$ 100 km s$^{-1}$ also tended to be those that were particularly blue for their spectral types and yielded older kinematic ages. Since neither source is particularly deviant in its colors, we conclude that both components are of field age, possibly tending toward the older range of field sources.  While there are differing results on what ``field age'' might mean for L and T dwarfs, recent results from \citet{Burgasser15} show that L dwarfs within $\sim$ 20 pc have kinematic mean ages of 6.5 $\pm$ 0.4 Gyr.   However we can only conclude that the L dwarf receives a field gravity using spectral indices and neither the L nor the T dwarf show features of a halo subdwarf.  Consequently we adopt a broad age range for this system; it appears to be older than 500 Myr and younger than $\sim$ 10 Gyr, a conservative cap on the age of field sources. 

\section{Fundamental parameters}
Following the prescription of \citet{Filippazzo15}, we use the Gaia DR2 parallax of W2150A combined with all available photometric and spectroscopic information to produce distance calibrated spectral energy distributions (SEDs) for the two sources.  By integrating over these SEDs, we directly calculate their bolometric luminosities. 
Using the evolutionary models of \citet{Saumon08} paired with the age range cited above, we obtain a radius range for each source and semi-empirically obtain estimates for the $T_{\rm eff}$, mass, and logg. For W2150A we used the Gaia G, 2MASS JHK, and WISE W1W2W3 photometry along with the optical and near infrared spectra to extract information.  For W2150B we only have the WISE W1W2 bands and {\it Spitzer} $ch1ch2$ bands along with the FIRE prism spectrum.  The \citet{Filippazzo15} method requires at least a near infrared estimate of photometry to scale the spectrum, therefore we estimated the 2MASS $J$ and $H$ bands using offsets computed from WISE and 2MASS detected T8 objects in \citet{Kirkpatrick11} (see Table ~\ref{tab:param} for details).  

We find that W2150A has log(L$_{bol}$/L$_{\sun}$)=-3.69 $\pm$ 0.01,  $T_{\rm eff}$=2118 $\pm$ 62 K, logg=5.2 $\pm$ 0,2, radius=1.03 $\pm$ 0.06R$_{Jup}$ and an estimated mass=72 $\pm$ 12 $M_{\rm Jup}$ while the T8 has log(L$_{bol}$/L$_{\sun}$)=-5.64 $\pm$ 0.02, $T_{\rm eff}$=719 $\pm$ 61 K, logg=4.9 $\pm$ 0,5, radius=0.95 $\pm$ 0.16 R$_{Jup}$ and an estimated mass=34 $\pm$ 22 $M_{\rm Jup}$. All of these fundamental parameter values are listed in Table ~\ref{tab:param}; they appear consistent with field sources discussed in \citet{Filippazzo15}.

\begin{figure}[ht!]
\begin{center}
\includegraphics[width=1.0\linewidth]{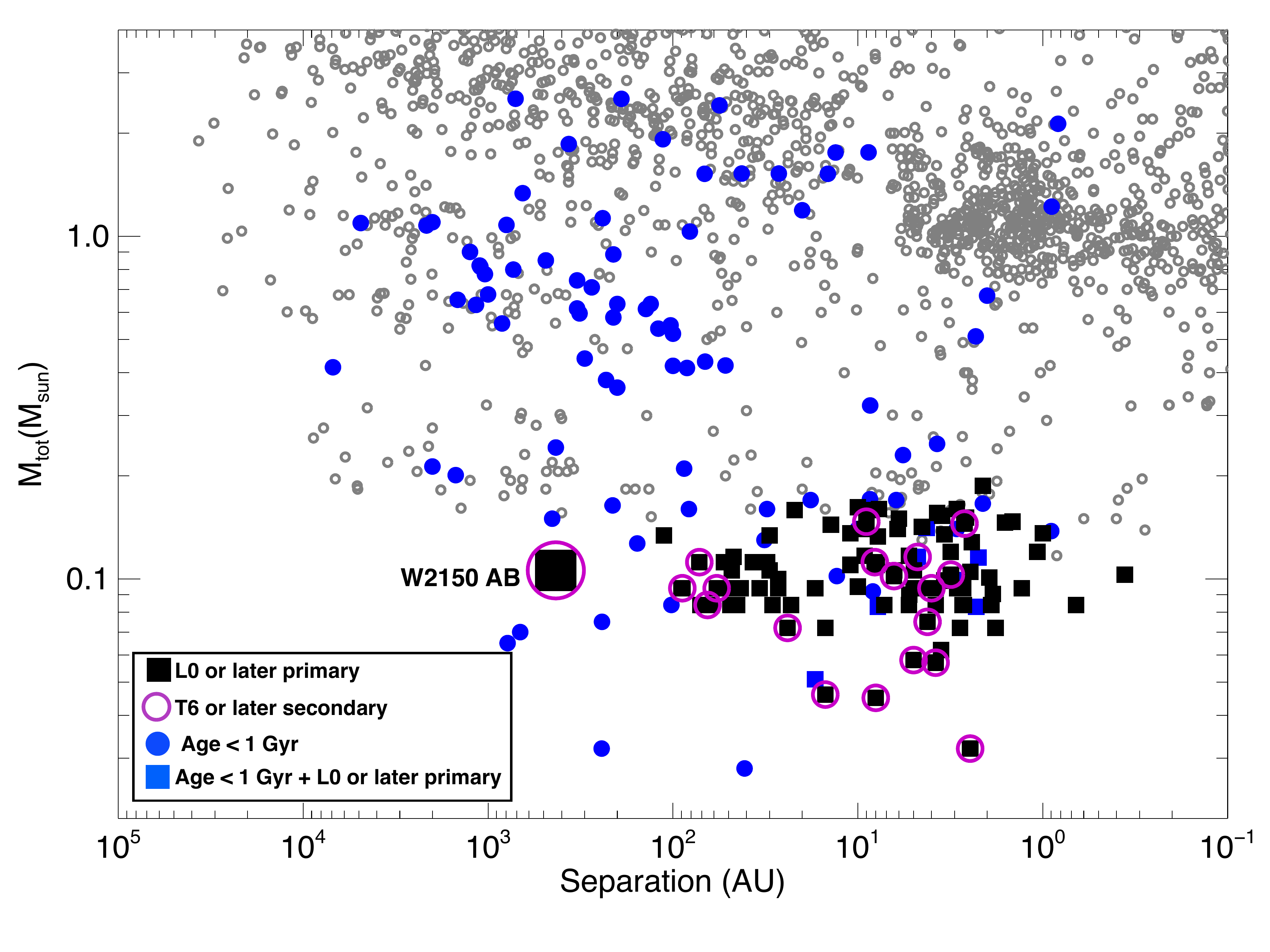}
\includegraphics[width=1.0\linewidth]{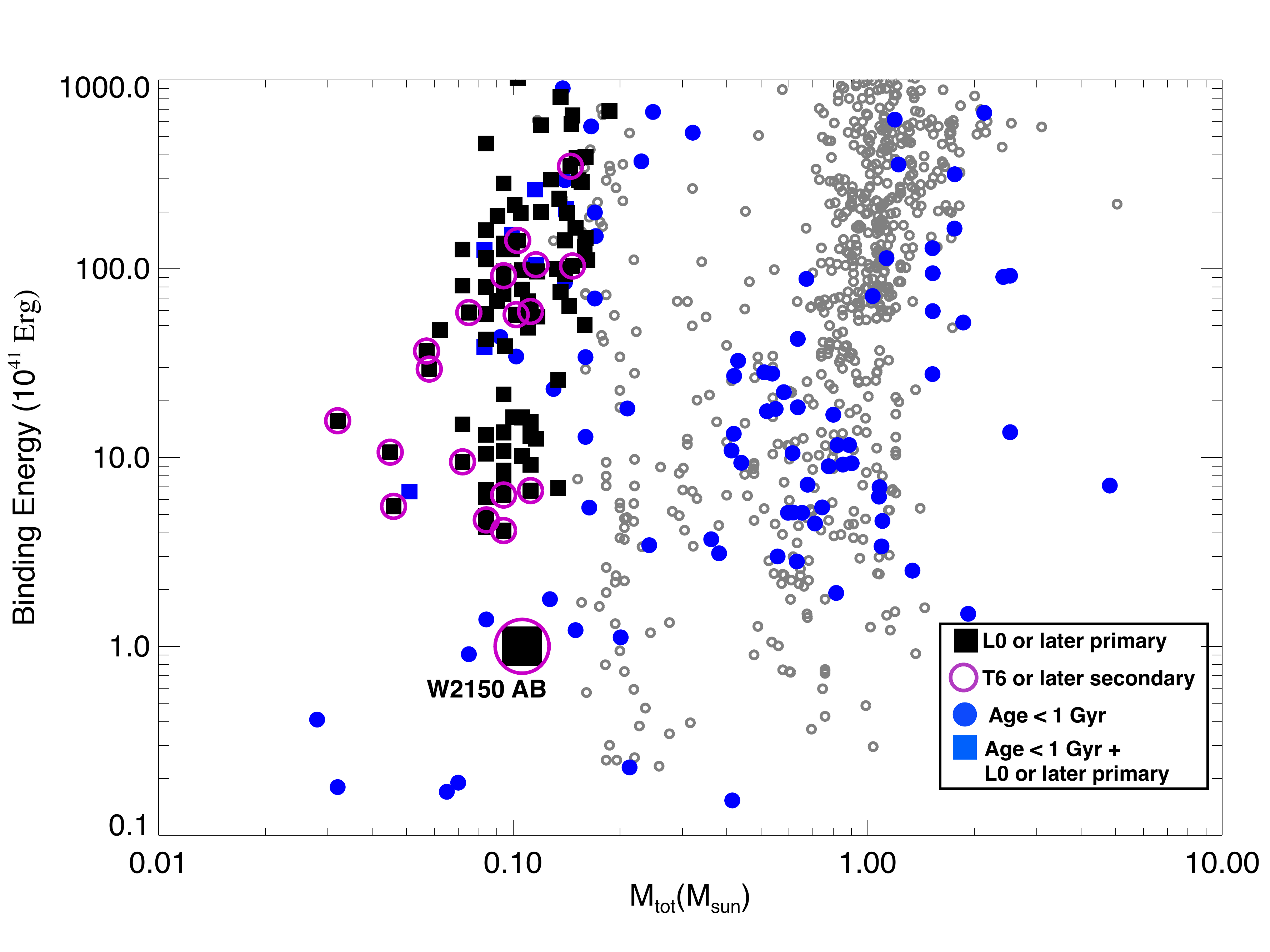}
\caption{A collection of systems from the literature representing co-moving or binary companions.  Systems with ages $<$ 1 Gyr are plotted as blue and field age objects are either light grey or black.  If the system contains at least an L dwarf or later primary, we plotted the system as a black square.  If it also contained a T6 or later secondary we plotted the system with an open purple circle.  We show total mass versus separation on the top panel and total mass versus binding energy on the lower panel. W2150AB stands out as the lowest binding energy system not in a young cluster.}
\label{fig:BindingEnergy1}
\end{center}
\end{figure}

\section{Binding Energy\label{sec:bindingenergy}}
\begin{figure}[ht!]
\begin{center}
\includegraphics[width=1.0\linewidth]{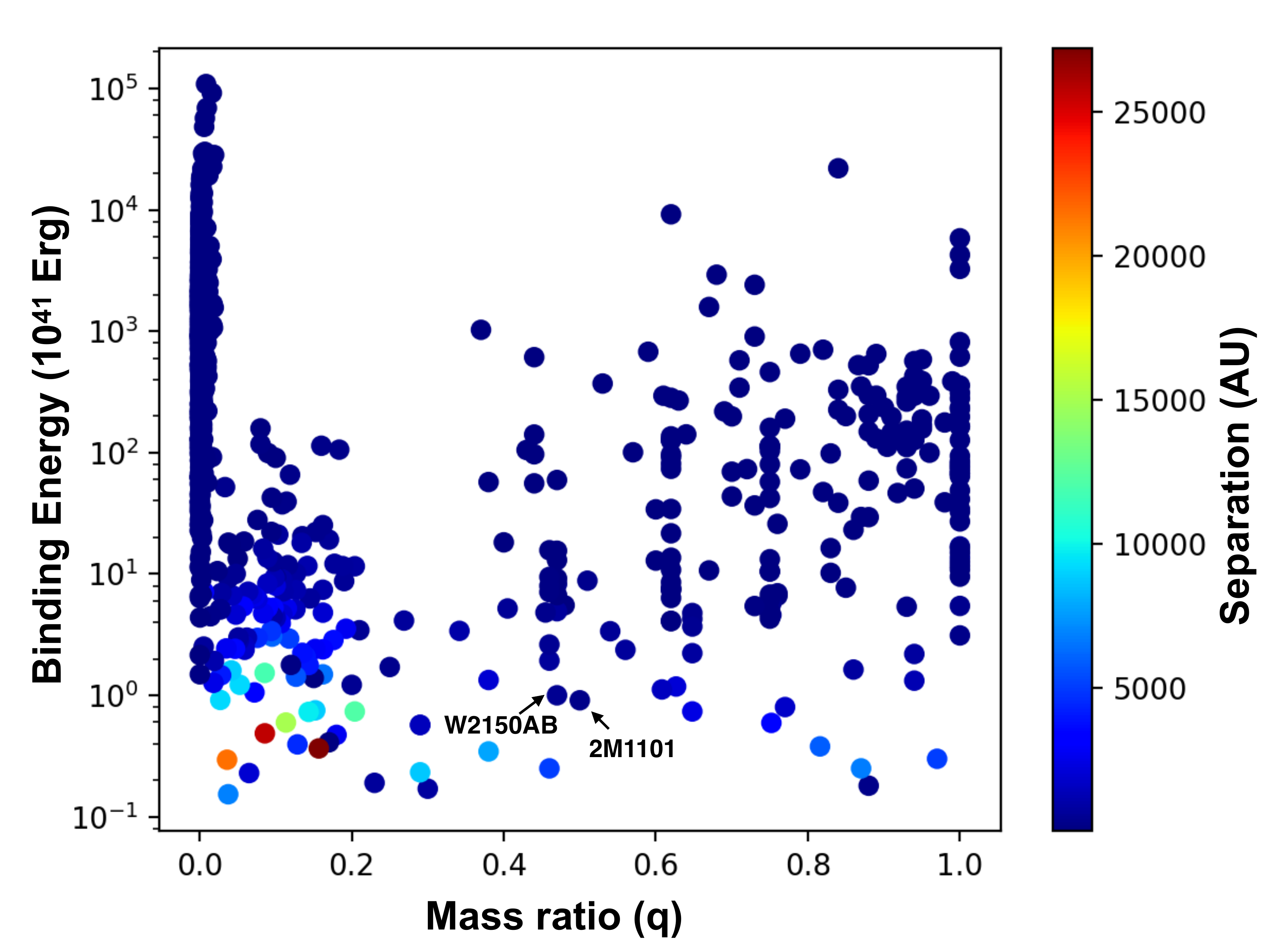}
\caption{The same systems plotted as in Figure ~\ref{fig:BindingEnergy1}. In this case we show the mass ratio (q) versus binding energy and color code the systems by their physical separation.  
\label{fig:BindingEnergy2}}
\end{center}
\end{figure}

Early studies of very low mass binaries concluded that there was a minimum binding energy (E$_{b}$) required for systems to form and remain stable.  Works such as \citet{Close03} and \citet{Burgasser03B} determined E$_{b}$ for field ultracool dwarf to ultracool dwarf pairs of $\sim$ 2 x 10$^{42}$ erg. However, as \citet{Faherty10} have discussed, searches in recent years have revealed numerous field-age and young wide companions falling well below the previously considered minimum binding energy (e.g. \citealt{Radigan09}, \citealt{Artigau09}, \citealt{Luhman04}, \citealt{Burgasser07}).  We attempted to compile an up-to-date sample of very low mass binaries/companions, drawing from both large survey papers such as \citet{Deacon14}, spectral binary papers such as \citet{Bardalez14}, and individual young companion papers such as \citet{Artigau15}, and \citet{Naud14}.
Figures~\ref{fig:BindingEnergy1} and  \ref{fig:BindingEnergy2} display a compilation of binary (or co-moving) systems from various catalogs. We list the objects displayed in Figures~\ref{fig:BindingEnergy1} and \ref{fig:BindingEnergy2} with M$_{tot}$ $<$ 0.2 M$_{Sun}$ organized by increasing binding energy in Table ~\ref{tab:companions}. Additionally we searched the Gaia DR2 catalog to see if either the primary or secondary component had astrometry and/or photometry reported and we list the results in Table ~\ref{tab:Gaiadr2companions}.  

We separate the different sub-populations of companion systems thought to be younger than 1 Gyr and those older on Figure~\ref{fig:BindingEnergy1}. The top panel of Figure~\ref{fig:BindingEnergy1} shows separation versus total mass of the system and allows us to examine if there is a distinguishable distance which would delineate where systems become unstable and disperse.  The bottom panel of Figure~\ref{fig:BindingEnergy2} shows the total mass of the system versus binding energy allowing us to investigate if there is a minimum value required for formation or survival within the Galaxy.  We find that W2150AB occupies a unique space on the upper and lower panels of Figure ~\ref{fig:BindingEnergy1}.  It is surrounded by only young sources with comparable low binding energy objects in the field with higher total masses.  It is the only L plus T dwarf co-moving system with a separation larger than $\sim$100 AU and it is one of only three systems where both the L and T dwarfs are resolved at all -- the others are Luhman 16AB (\citealt{Luhman13}) and SDSS J1416+13AB (\citealt{Burningham10}).

Assuming that W2150 is field age (see section ~\ref{sec:age} above), we can find an analog to its mass ratio and binding energy among the young systems ($<$ 1 Gyr) compiled.  For instance, the $\sim$ 2 Myr Chamaeleon star forming region contains 2MASS J11011926-7732383AB (2M1101AB), an M7.25 and an M8.25 with a mass ratio of $\sim$0.5 and a separation of $\sim$ 240 AU (\citealt{Luhman04}).  As can be seen in Figure ~\ref{fig:BindingEnergy2}, 2M1101AB has a similar mass ratio and binding energy to W2150AB.  At the time of its discovery, 2M1101AB was the first brown dwarf binary discovered with a separation $>$ 20 AU and its existence was celebrated as a definitive insight into the formation of brown dwarfs.  W2150AB now shows that such systems can survive into the field.  

While it is intriguing to see the two systems with similar properties, they differ in age by several Gyr.  Moreover, 2M1101AB is a member of Chamaeleon which is a densely crowded area of star formation compared to the sparsely-separated field where we find W2150AB.  While we have no idea how or where W2150AB formed, we can investigate the feasibility that it could have formed like 2M1101AB and survived dynamical interactions in its natal cluster and the Galaxy until a field age. As was discussed in \citealt{Burgasser03B} and adapted in \citet{Close07}, systems with total masses $\sim$ 0.1 M$_{\sun}$ will be stable against any type of stellar encounter as long as the separation is $<$ 1800 AU and the number density of stellar perturbers is on the order of the measured Galactic disk mass density (see \citealt{Pham97}, \citealt{Holmberg00}). W2150AB meets these criteria. Consequently, it is not altogether surprising that W2150AB exists -- even though no other system had been discovered to date that rivals its properties in the field.  What is surprising and what has been discussed at length concerning the difference between young systems discovered and those in the field, is whether very low mass companions born in star forming regions should become unstable and evaporate over time given their more dense natal environments.  \citealt{Burgasser03B} postulated how the local stellar density at birth impacts the viability of a pair surviving.  \citet{Close07} built upon that work and examined the stellar density for clusters such as Chamaeleon, Rho Ophiucus, Upper Scorpius, etc., where the bulk of the young sources from Figure ~\ref{fig:BindingEnergy1} have been discovered.  Consequently, \citet{Close07} suggested that the majority of young systems (including 2M1101AB) were in the process of evaporating.  Following this logic, W2150AB could either be considered a system that formed in a low density environment (small sparse cluster for instance) or, it was perturbed into its wide configuration while in a more dense cluster but the environment dispersed before it could evaporate the binary, leaving the system stable in the less dense field environment.
 
 Another possible explanation for the survival of this source to field ages could be that one of their components is in reality a closely-separated binary system, making this a hierarchical triple system while increasing the overall binding energy. To test the latter hypothesis, we followed the prescription of \citet{Burgasser10} to identify spectral binary systems from near-infrared, low-resolution, SpeX spectra. However, we do not find peculiarities in the spectrum of the primary attributable to unresolved binarity.  Unfortunately this technique is not applicable for T8 objects therefore we can not rule out that the secondary might harbor an unseen companion.
 
Regardless of the reason it has survived until today, the total estimated mass of the system is 0.106 M$_{Sun}$. So with a physical separation as large as 341 AU, the binding energy (E$_{bin}$=10$^{41}$ erg) of W2150AB is the lowest found among ultracool dwarf objects not identified in a young cluster.

\section{Discussion}
We report the discovery of a resolved L1+T8 co-moving system: W2150AB with a physical separation of $\sim$ 341 AU.  This discovery was enabled by a dedicated cohort of citizen scientists participating in the Backyard Worlds: Planet 9 citizen science project.  The cool secondary appeared in no online catalogs, so it had eluded astronomers performing automated searches. 

We obtained Magellan FIRE prism near infrared spectra for both the primary and secondary and find both sources appear comparable to field sources with no deviant or peculiar features. The primary in the system is also a Gaia detected source and has a well determined parallax of 41.3593 $\pm$ 0.2799 mas and proper motion components of ($\mu_{ra}$cos(dec), $\mu_{dec}$)=(888.627 $\pm$ 0.502, -298.234 $\pm$ 0.518) mas.  Assuming an age range for the system of 0.5 - 10 Gyr we find that W2150A has log(L$_{bol}$/L$_{\sun}$)=-3.69 $\pm$ 0.01,  $T_{\rm eff}$=2118 $\pm$ 62 K, logg=5.2 $\pm$ 0,2, radius=1.03 $\pm$ 0.06R$_{Jup}$ and an estimated mass=72 $\pm$ 12 $M_{\rm Jup}$ while W2150B has log(L$_{bol}$/L$_{\sun}$)=-5.64 $\pm$ 0.02, $T_{\rm eff}$=719 $\pm$ 61 K, logg=4.9 $\pm$ 0,5, radius=0.95 $\pm$ 0.16 R$_{Jup}$ and an estimated mass=34$\pm$ 22 $M_{\rm Jup}$.  The total estimated mass of the system is 0.106 M$_{Sun}$ hence with a physical separation as large as 341 AU, the binding energy (E$_{bin}$=10$^{41}$ erg) is the lowest found among ultracool dwarf objects not identified in a young cluster.  In separation, E$_{bin}$, and mass ratio, W2150AB resembles 2M1101AB, the first brown dwarf binary discovered with a separation $>$ 20 AU. 2M1101AB, discovered in the Chamaeleon star forming region, was heralded as a source of definitive insight into the formation of brown dwarfs. But W2150AB leaves us with an intriguing question about whether it is an evolved version of 2M1101AB or perhaps a system that formed in a low density cluster that survived unperturbed by interactions with nearby stellar or giant molecular clouds. Given that it is easily resolved with ground or space based observatories, W2150AB is an excellent benchmark system for understanding how brown dwarfs form and evolve together.

\acknowledgments
The Backyard Worlds: Planet 9 team would like to thank the many Zooniverse volunteers who have participated in this project,
from providing feedback during the beta review stage to classifying flipbooks to contributing to the discussions on TALK. We
would also like to thank the Zooniverse web development team for their work creating and maintaining the Zooniverse platform
and the Project Builder tools. This research was supported by NASA Astrophysics Data Analysis Program grant NNH17AE75I.

This research has made use of: the Washington Double Star Catalog maintained at the U.S. Naval Observatory; the SIMBAD database and VizieR catalog access tool, operated at the Centre de Donnees astronomiques de Strasbourg, France (\citealt{Ochsenbein00}); data products from the Two Micron All Sky Survey (2MASS; \citealt{Skrutskie06}), which is a joint project of the University of Massachusetts and the Infrared Processing and Analysis Center (IPAC)/California Institute of Technology (Caltech), funded by the National Aeronautics and Space Administration (NASA) and the National Science Foundation; data products from the Wide-field Infrared Survey Explorer (WISE; and \citealt{Wright10}), which is a joint project of the University of California, Los Angeles, and the Jet Propulsion Laboratory (JPL)/Caltech, funded by NASA.  This project was developed in part at the 2017 Heidelberg Gaia Sprint, hosted by the Max-Planck-Institut fur Astronomie, Heidelberg. AMM acknowledges support from Hubble Fellowship HST-HF2-51415.001-A and NASA ADAP grant NNH17AE75I. SEL is supported by an appointment to the NASA Postdoctoral Program at NASA Goddard Space Flight Center, administered by Universities Space Research Association under contract with NASA. This work has made use of data from the European Space Agency(ESA) mission Gaia, processed by the Gaia Data Processing and Analysis Consortium. Funding for the DPAC has been provided by national institutions, in particular the institutions participating in the Gaia Multilateral Agreement.

\clearpage
\startlongtable
\begin{deluxetable*}{cccccc}
\tablecaption{Measured Parameters\label{tab:param}}
\tablehead{
\colhead{Parameter} & \colhead{W2150A} & \colhead{W2150B} & \colhead{System} & \colhead{Units} & \colhead{Reference}}
\colnumbers
\startdata
{\bf ASTROMETRY}\\
\hline
\hline
$\alpha$ & 327.58083685735\tablenotemark{a} ($\pm$0.2\,mas) & 327.579158\tablenotemark{e} ($\pm$0.6\,mas)& $\cdots$ & deg & 1,2 \\
$\delta$ & -75.34482002123\tablenotemark{a} ($\pm$0.3\,mas) & -75.348499\tablenotemark{e} ($\pm$0.6\,mas)&  $\cdots$ & deg & 1,2 \\
$\ell$\tablenotemark{a}    & 315.9126	& $\cdots$ &$\cdots$ &deg & 1 \\
$b$\tablenotemark{a}       & -36.8406  & $\cdots$ &$\cdots$ &deg & 1\\
$\varpi$       & 41.3593\,$\pm$\,0.2799 &$\cdots$ &$\cdots$ & mas & 1\\
$\mu_{\alpha}$ & 888.627\,$\pm$\,0.502 & 876\,$\pm$\,45 & $\cdots$ & mas\,yr$^{-1}$ & 1,2\\
$\mu_{\delta}$ & -298.234\,$\pm$\,0.518  & -278\,$\pm$\,45 & $\cdots$ & mas\,yr$^{-1}$ & 1,2\\
\hline
\hline
{\bf PHOTOMETRY}\\
\hline
\hline
$G_{BP}$ & 21.4462$\pm$0.2021 & $\cdots$ & $\cdots$ & mag &  1\\
$G$     & 18.9110$\pm$0.0038 & $\cdots$ & $\cdots$ & mag &  1\\
$G_{RP}$ & 17.3031$\pm$0.0118 & $\cdots$ & $\cdots$ & mag &  1\\
$I$ & 17.53$\pm$0.17 & $\cdots$ & $\cdots$ & mag &  5\\
$J$ & 14.056$\pm$0.029  & (19.03$\pm$0.23)\tablenotemark{f} & $\cdots$ &  mag &3\\
$H$ & 13.176$\pm$0.032  & (19.24$\pm$0.22)\tablenotemark{f} & $\cdots$ &  mag &3\\
$K_s$ & 12.673$\pm$0.030 & $\cdots$ & $\cdots$ &  mag &3\\
$W1$\tablenotemark{b} & 12.317$\pm$0.024  & 18.18$\pm$0.15 & $\cdots$ & mag &4,2\\
$W2$\tablenotemark{b} & 12.053$\pm$0.023  & 15.81$\pm$0.05 & $\cdots$ & mag &4,2\\
$W3$\tablenotemark{b}& 11.616$\pm$0.150 & $\cdots$ & $\cdots$ & mag &4\\
$W4$\tablenotemark{b} & $<$9.328          &  $\cdots$        & $\cdots$ & mag &4\\
$ch1$& $\cdots$ & 17.01$\pm$0.03   & $\cdots$ &   mag &2\\
$ch2$& $\cdots$ & 15.60$\pm$0.02   &$\cdots$ &   mag &2\\
\hline
\hline
{\bf SPECTROSCOPY}\\
\hline
\hline
Spectral Type (OpT) & L1$\pm$1 & $\cdots$ & $\cdots$ & $\cdots$ &2\\
Spectral Type (IR) & L1$\pm$1 & T8$\pm$1 & $\cdots$ & $\cdots$ &2\\
\hline
\hline
{\bf FUNDAMENTALS}\\
\hline
\hline
Age & 0.5 - 10 & 0.5 -10 & 0.5 - 10 & Gyr & 2\\
log($L_{\rm bol}$/$L_{\odot}$) & -3.69$\pm$0.01 & -5.64 $\pm$ 0.02 & $\cdots$ & $\cdots$ &2\\
$T_{\rm eff}$ & 2118$\pm$62 & 719$\pm$61 & $\cdots$ & K &2\\
Radius        & 1.03$\pm$0.06 & 0.95$\pm$0.16&$\cdots$ & $R_{\rm Jup}$ & 2\\
Mass          & 72$\pm$12 & 34$\pm$22 & $\cdots$  & $M_{\rm Jup}$&2\\
$\log{g}$        & 5.2$\pm$0.2 & 4.9$\pm$0.5 & $\cdots$  & $\cdots$ & 2\\
\hline
\hline
{\bf KINEMATICS}\\
\hline
\hline
Distance\tablenotemark{c}       & 24.18$\pm$0.16 & 27$\pm$4\tablenotemark{g} & 24.18$\pm$0.16 & pc & 2\\
$v_{\rm tan}$\tablenotemark{d} & 107.43\,$\pm$\,0.06 &117$\pm$6 & 107.43\,$\pm$\,0.06 & km\,s$^{-1}$ & 2\\
\hline
\hline
{\bf ABS MAGS}\\
\hline
\hline
M$_{G}$    & 16.99$\pm$ 0.02&$\cdots$&$\cdots$& mag & 2\\
M$_{I}$    & 15.61$\pm$ 0.17&$\cdots$&$\cdots$& mag & 2\\
M$_{J}$  & 12.14$\pm$0.03&$\cdots$&$\cdots$& mag & 2\\
M$_{H}$  & 11.26$\pm$ 0.03&$\cdots$&$\cdots$& mag & 2\\
M$_{K}$  & 10.76$\pm$ 0.03&$\cdots$&$\cdots$& mag & 2\\
M$_{W1}$ & 10.43$\pm$ 0.03& 16.26$\pm$0.15 & $\cdots$ & mag & 2\\
M$_{W2}$ & 10.16$\pm$ 0.03& 13.89$\pm$ 0.05 &$\cdots$ & mag & 2\\
M$_{W3}$ & 9.55$\pm$ 0.16&$\cdots$&$\cdots$& mag & 2\\
M$_{ch1}$ & $\cdots$& 15.09$\pm$0.03 & $\cdots$ & mag & 2\\
M$_{ch2}$ & $\cdots$& 13.68$\pm$0.03 & $\cdots$ & mag & 2\\
\hline
\hline
{\bf SYSTEM}\\
\hline
\hline
Separation&$\cdots$&$\cdots$&14.1 & $\arcsec$ & 2\\
Separation&$\cdots$&$\cdots$&341 & AU & 2\\
Binding Energy&$\cdots$ & $\cdots$ &1.004 & 10 $^{41}$ erg  & 2\\
\enddata
\tablenotetext{a}{epoch J2015.5, ICRS}
\tablenotetext{b}{For the L1 primary, we chose the original \emph{WISE} catalog values in the analysis over the AllWISE values so we could compare to the photometry in \citet{Faherty16},  For the T8 secondary W2 comes from the yearly W2 coadd analysis of $\S$\ref{sec:secondary}, and W1 comes from the unWISE Catalog \citep{Schlafly19} with a correction applied for source motion.}
\tablenotetext{c}{Calculated using D = 1/$\pi$, which is good approximation for parallax known to $\pi$/$\sigma_{\pi}$ = 133 accuracy}
\tablenotetext{d}{Calculated using \citet{GaiaDR2} astrometry.}
\tablenotetext{e}{Calculated using WISE image at MJD=57947.}
\tablenotetext{f}{The 2MASS J and H magnitudes are estimates from the expected J-W2 and H-W2 colors of a T8.  We used the sample of brown dwarfs in \citet{Kirkpatrick11} to estimate an offset of 3.21$\pm$0.23 mag for 2MASS J and 3.43$\pm$0.23 mag for 2MASS H from the WISE W2 magnitude.}
\tablenotetext{g}{Calculated using the  M$_{W2}$ from the SpT relation in \citet{Kirkpatrick19}}
\tablecomments{The object does not have entries in GSC 2.2, USNO-B1.0, and does not appear on any of the photographic sky surveys scanned by SuperCOSMOS. \\
References: (1) \citet{GaiaDR2}, (2) This paper, (3) \citet{Cutri03}, (4) \citet{Wright10}.}, (5) \citet{DENIS05}
\end{deluxetable*}

\input{Table1-Sub.tex}
\input{Table2-Sub.tex}

\vspace{5mm}
\facilities{\emph{Gaia}, Hale(TripleSpec), \emph{WISE}, CTIO:2MASS, \emph{UKIRT}}

\software{Aladin, BANYAN~$\Sigma$ (\citealt{Gagne18})}

\bibliography{GaiaLdwarf.bib}
\end{document}

%% file: Table1-Sub.tex
\rotate
\tabletypesize{\scriptsize}
\clearpage
\begin{deluxetable*}{lcccccccccccc}
\tablewidth{0pt}
\tablecolumns{15}
\tablecaption{Companion systems with total mass $<$ 0.2 $M_{Sun}$\label{tab:companions}\tablenotemark{d}}
\tablewidth{0pt}
\tablehead{
\colhead{Name\tablenotemark{h}} &
\colhead{RA\tablenotemark{i}} &
\colhead{DEC\tablenotemark{i}} &
\colhead{SpT$_{M1}$} &
\colhead{SpT$_{M2}$} &
\colhead{Mass$_{M1}$\tablenotemark{e}} &
\colhead{Mass$_{M2}$\tablenotemark{e}} &
\colhead{M$_{tot}$}&
\colhead{(q)} &
\colhead{Sep}&
\colhead{E$_{bin}$\tablenotemark{f}} &
\colhead{Young?\tablenotemark{g}}&
\colhead{Reference}\\
&
&
&
&
&
\colhead{M$_{Sun}$}&
\colhead{M$_{Sun}$}&
\colhead{M$_{Sun}$}&
&
\colhead{AU}&
\colhead{x10$^{41}$ erg}\\
\colhead{(1)}&
\colhead{(2)}&
\colhead{(3)}&
\colhead{(4)}&
\colhead{(5)}&
\colhead{(6)}&
\colhead{(7)}&
\colhead{(8)}&
\colhead{(9)}&
\colhead{(10)}&
\colhead{(11)}&
\colhead{(12)}&
\colhead{(13)}\\
}
\startdata
FUTau	&	04 23 35.38	&	+25 03 03.0	&	M7.25	&	M9.25	&	0.05	&	0.015	&	0.065	&	0.30	&	784	&	0.17	&	Yng	&	68	\\
Oph-11	&	16 22 25.20	&	-24 05 14.0	&	M9	&	M9.5	&	0.017	&	0.015	&	0.032	&	0.88	&	243	&	0.18	&	Yng	&	49	\\
UScoCTIO-108	&	16 05 53.94	&	-18 18 42.7	&	M7	&	M9.5	&	0.057	&	0.013	&	0.07	&	0.23	&	670	&	0.19	&	Yng	&	10	\\
1258+4013	&	12 58 35.01	&	+40 13 08.3	&	M6	&	M7	&	0.105	&	0.091	&	0.196	&	0.87	&	6700	&	0.25	&	Not Yng	&	79	\\
NLTT730	&	00 15 02.40	&	+29 59 29.8	&	M4	&	L7.5	&	0.125	&	0.058	&	0.183	&	0.46	&	5070	&	0.25	&	Not Yng	&	37	\\
0126-5022	&	01 26 55.50	&	-50 22 39.0	&	M6.5	&	M8	&	0.095	&	0.092	&	0.187	&	0.97	&	5100	&	0.30	&	Not Yng	&	3	\\
2M1207	&	12 07 33.40	&	-39 32 54.0	&	M8	&	L5	&	0.024	&	0.004	&	0.028	&	0.17	&	41	&	0.41	&	Yng	&	29	\\
LHS6176	&	09 50 47.28	&	+01 17 34.3	&	M4	&	T8	&	0.125	&	0.036	&	0.161	&	0.29	&	1400	&	0.57	&	Not Yng	&	69	\\
LHS2803	&	13 48 02.90	&	-13 44 07.1	&	M4.5	&	T5	&	0.125	&	0.036	&	0.161	&	0.29	&	1400	&	0.57	&	Not Yng	&	36	\\
Koenigstuhl-1	&	00 21 05.91	&	-42 44 43.5	&	M6	&	M9.5	&	0.103	&	0.079	&	0.182	&	0.77	&	1800	&	0.80	&	Not Yng	&	28	\\
1101-7732	&	11 01 19.26	&	-77 32 38.3	&	M7	&	M8	&	0.05	&	0.025	&	0.075	&	0.50	&	242	&	0.91	&	Yng	&	67	\\
2150-7520	&	21 50 18.99	&	-75 20 54.6	&	L1	&	T8	&	0.072	&	0.034	&	0.106	&	0.47	&	430	&	1.00	&	Not Yng	&	96	\\
LP261-75	&	09 51 05.49	&	+35 58 02.1	&	M4.5	&	L6	&	0.125	&	0.025	&	0.15	&	0.20	&	450	&	1.22	&	Yng	&	83	\\
1328+0808	&	13 28 35.49	&	+08 08 19.5	&	M6	&	M8.5	&	0.1	&	0.094	&	0.194	&	0.94	&	1250	&	1.32	&	Not Yng	&	97	\\
VHS1256	&	12 56 01.92	&	-12 57 23.9	&	M7.5	&	L7	&	0.073	&	0.011	&	0.084	&	0.15	&	102	&	1.39	&	Yng	&	42	\\\enddata 
\tablenotetext{a}{Dynamical masses from \citet{Dupuy17}}
\tablenotetext{b}{Total masses listed in \citet{Dupuy17} but individual masses were not therefore we used the relation in that work to estimate object masses.}
\tablenotetext{c}{Dynamical masses from \citet{Garcia17}}
\tablenotetext{d}{Dynamical masses from \citet{Dieterich18}}
\tablenotetext{e}{Masses were either obtained from dynamical mass measurements (and noted as such), discovery paper estimates or by applying the spectral type to mass relation from \citet{Dupuy17}}
\tablenotetext{f}{To account for a distribution of the eccentricities of the binary orbits, we multiply the physical separation by 1.26 and use this value to compute the binding energy.}
\tablenotetext{g}{Yng indicates a source is associated with a cluster/moving-group/star forming region with an age $<$ 1 Gyr.  Not Yng indicates no associated region therefore a field age for the source.}
\tablenotemark{h}{The name used is a shorthand from the discovery paper or from the coordinates of the primary.}
\tablenotemark{i}{The RA and DEC listed are of the primary in the system.}
\tablecomments{This stubtable is a preview of the entire sample, which will be available as a machine readable table.  All of Table 1 and Table 2 data can be found on a viewable and commentable google spreadsheet here: https://drive.google.com/file/d/15Eidlmsidj5kxqj2bc4HNrGdHJJZnjY0/view?usp=sharing }
\end{deluxetable*} 

%% file: Table2-Sub.tex
\rotate
\tabletypesize{\scriptsize}
\clearpage
\begin{deluxetable*}{lcccccccccccccc}
\tablewidth{0pt}
\tablecolumns{15}
\tablecaption{Companion systems with total mass $<$ 0.2 $M_{Sun}$\label{tab:Gaiadr2companions}\tablenotemark{d}}
\tablewidth{0pt}
\tablehead{
\colhead{Name} &
\colhead{RA} &
\colhead{DEC} &
\colhead{$\pi$}&
\colhead{$\mu_{\alpha}$} &
\colhead{$\mu_{\delta}$} &
\colhead{$G$} &
\colhead{RA} &
\colhead{DEC} &
\colhead{$\pi$}&
\colhead{$\mu_{\alpha}$} &
\colhead{$\mu_{\delta}$} &
\colhead{$G$}\\
&
\colhead{Prim}&
\colhead{Prim}&
\colhead{Prim}&
\colhead{Prim}&
\colhead{Prim}&
\colhead{Prim}&
\colhead{Sec}&
\colhead{Sec}&
\colhead{Sec}&
\colhead{Sec}&
\colhead{Sec}&
\colhead{Sec}\\
&
&
&
\colhead{mas}&
\colhead{mas yr$^{-1}$}&
\colhead{mas yr$^{-1}$}&
&
&
&
\colhead{mas}&
\colhead{mas yr$^{-1}$}&
\colhead{mas yr$^{-1}$}\\
\colhead{(1)}&
\colhead{(2)}&
\colhead{(3)}&
\colhead{(4)}&
\colhead{(5)}&
\colhead{(6)}&
\colhead{(7)}&
\colhead{(8)}&
\colhead{(9)}&
\colhead{(10)}&
\colhead{(11)}&
\colhead{(12)}&
\colhead{(13)}\\
}
\startdata
FUTau	&	04 23 35.39	&	+25 03 02.7	&	7.5981	$\pm$	0.1497	&	6.895	$\pm$	0.376	&	-21.026	$\pm$	0.202	&	15.24	&	04 23 35.73	&	+25 02 59.6	&	7.4909	$\pm$	1.2887	&	12.450	$\pm$	4.056	&	-21.761	$\pm$	1.903	&	20.48	\\
Oph-11	&	16 22 25.20	&	-24 05 13.6	&	7.3440	$\pm$	0.4060	&	-15.790	$\pm$	0.698	&	-23.236	$\pm$	0.520	&	18.92	&	16 22 25.19	&	-24 05 15.9	&	$\cdots$			&	$\cdots$			&	$\cdots$			&	20.24	\\
UScoCTIO-108	&	16 05 54.07	&	-18 18 44.3	&	6.9306	$\pm$	0.2409	&	-10.156	$\pm$	0.454	&	-20.618	$\pm$	0.260	&	17.42	&	$\cdots$	&	$\cdots$	&	$\cdots$			&	$\cdots$			&	$\cdots$			&	$\cdots$	\\
1258+4013	&	12 58 35.01	&	+40 13 08.1	&	8.4520	$\pm$	0.3382	&	79.289	$\pm$	0.356	&	-111.430	$\pm$	0.445	&	19.17	&	12 58 37.99	&	+40 14 01.5	&	7.3800	$\pm$	0.5788	&	77.913	$\pm$	0.539	&	-109.434	$\pm$	0.833	&	19.83	\\
NLTT730	&	00 15 05.73	&	+29 55 40.3	&	28.6060	$\pm$	0.0709	&	381.542	$\pm$	0.103	&	-227.076	$\pm$	0.077	&	14.11	&	$\cdots$	&	$\cdots$	&	$\cdots$			&	$\cdots$			&	$\cdots$			&	$\cdots$	\\
0126-5022	&	01 26 55.50	&	-50 22 38.7	&	13.9950	$\pm$	0.2440	&	140.207	$\pm$	0.333	&	-50.356	$\pm$	0.311	&	18.80	&	01 27 02.83	&	-50 23 20.9	&	14.4188	$\pm$	0.3546	&	146.684	$\pm$	0.442	&	-45.784	$\pm$	0.421	&	19.28	\\
2M1207	&	12 07 33.46	&	-39 32 54.0	&	15.5242	$\pm$	0.1561	&	-64.083	$\pm$	0.233	&	-23.720	$\pm$	0.130	&	17.41	&	$\cdots$	&	$\cdots$	&	$\cdots$			&	$\cdots$			&	$\cdots$			&	$\cdots$	\\
LHS6176	&	09 50 49.59	&	+01 18 13.6	&	50.8011	$\pm$	0.0762	&	234.644	$\pm$	0.109	&	-360.553	$\pm$	0.096	&	12.63	&	$\cdots$	&	$\cdots$	&	$\cdots$			&	$\cdots$			&	$\cdots$			&	$\cdots$	\\
LHS2803	&	13 48 07.27	&	-13 44 31.5	&	54.9974	$\pm$	0.0838	&	-687.597	$\pm$	0.144	&	-512.979	$\pm$	0.124	&	13.54	&	$\cdots$	&	$\cdots$	&	$\cdots$			&	$\cdots$			&	$\cdots$			&	$\cdots$	\\
Koenigstuhl-1	&	00 21 10.74	&	-42 45 40.1	&	37.3985	$\pm$	0.0727	&	255.046	$\pm$	0.088	&	-12.530	$\pm$	0.092	&	15.37	&	00 21 05.91	&	-42 44 43.4	&	38.5223	$\pm$	0.5439	&	258.636	$\pm$	0.733	&	-1.884	$\pm$	0.702	&	18.35	\\
1101-7732	&	11 01 19.19	&	-77 32 38.7	&	5.4081	$\pm$	0.1877	&	-22.653	$\pm$	0.435	&	2.062	$\pm$	0.397	&	18.33	&	11 01 19.42	&	-77 32 37.5	&	5.4333	$\pm$	0.3368	&	-23.668	$\pm$	0.748	&	1.931	$\pm$	0.723	&	19.40	\\
W2150	&	21 50 15.77	&	-75 20 36.7	&	41.3593	$\pm$	0.2799	&	888.627	$\pm$	0.502	&	-298.234	$\pm$	0.518	&	18.91	&	$\cdots$	&	$\cdots$	&	$\cdots$			&	$\cdots$			&	$\cdots$			&	$\cdots$	\\
LP261-75	&	09 51 04.44	&	 35 58 06.7	&	29.4464	$\pm$	0.1376	&	-100.975	$\pm$	0.142	&	-171.822	$\pm$	0.126	&	13.83	&	$\cdots$	&	$\cdots$	&	$\cdots$			&	$\cdots$			&	$\cdots$			&	$\cdots$	\\
1328+0808	&	13 28 35.39	&	 08 08 18.8	&	9.2012	$\pm$	0.3601	&	-148.679	$\pm$	0.718	&	-57.100	$\pm$	0.418	&	19.07	&	$\cdots$	&	$\cdots$	&	$\cdots$			&	$\cdots$			&	$\cdots$			&	$\cdots$	\\
VHS1256	&	12 56 01.84	&	-12 57 24.8	&	$\cdots$			&	$\cdots$			&	$\cdots$			&	15.05	&	$\cdots$	&	$\cdots$	&	$\cdots$			&	$\cdots$			&	$\cdots$			&	$\cdots$	\\\enddata 
\tablecomments{This stubtable is a preview of the entire sample, which will be available as a machine readable table. All of Table 1 and Table 2 data can be found on a viewable and commentable google spreadsheet here: https://drive.google.com/file/d/15Eidlmsidj5kxqj2bc4HNrGdHJJZnjY0/view?usp=sharing }
\end{deluxetable*}